\documentclass[twocolumn,aps,superscriptaddress]{revtex4-2}
\usepackage{amsmath, amssymb, amsthm, graphicx, epsfig, fancyhdr,epsfig,multirow}
\usepackage[utf8]{inputenc}
\usepackage{amsmath}
\usepackage{amsfonts}
\usepackage{amssymb}
\usepackage{xcolor}
\usepackage{comment}
\usepackage[normalem]{ulem}
\usepackage{tabularx}
\usepackage{comment}
\usepackage{adjustbox}
\usepackage{diagbox}
\usepackage[left=2cm,right=2cm,top=2cm,bottom=2cm]{geometry}
\usepackage {ulem}

\usepackage{color}

\renewcommand{\sout}{\bgroup \color{red} \ULdepth=-.5ex \ULset}

\usepackage{xcolor}

\usepackage{lineno}

\begin{document}
\title{Observation of Strong $\phi$-meson Directed Flow at High Baryon Density}
\author{The STAR Collaboration}

\begin{abstract}

We report the first observation of a large directed flow ($v_1$) of $\phi$ mesons in mid-central Au+Au collisions at $\sqrt{s_{\mathrm{NN}}}=3.0$--4.5 GeV, together with the $\sqrt{s_{\mathrm{NN}}}=7.7$ GeV collider-energy result for the collision-energy dependence, measured by the STAR experiment at RHIC. 
Despite its mesonic nature, the $\phi$-meson $v_{1}$ exhibits a magnitude comparable to that of protons and $\Lambda$ baryons, and is significantly larger (by a factor of $\gtrsim3$) than that of $K^0_S$ mesons at the lowest energies. The midrapidity slope $dv_1/dy$ shows a pronounced energy dependence similar to that observed for protons and $\Lambda$ baryons, in contrast to lighter mesons, in the high-baryon-density region. Comparisons with hadronic transport calculations indicate that the observed $\phi$-meson directed flow is consistent with production via high-mass baryon resonances and the collective motion of baryons influenced by baryonic mean-field interactions. Due to its long lifetime and relatively weak effective $\phi$--$N$ interaction, the $\phi$ meson can retain sensitivity to collective dynamics established during the dense baryonic stage of the collision. These results provide new experimental constraints on baryon-resonance production and strange-hadron transport dynamics in dense QCD matter.

\end{abstract}
\maketitle


Relativistic heavy-ion collisions aim to explore the phase structure of strongly interacting matter governed by Quantum Chromodynamics (QCD) under extreme temperature and baryon density conditions. Anisotropic flow, particularly the directed flow ($v_1$), the first harmonic coefficient, is a key observable that is sensitive to the early-time dynamics and the equation of state (EoS) of the created medium \cite{Stoecker:1986ci,Brachmann:1999xt}. In the RHIC Beam Energy Scan Phase-I (BES-I), the STAR experiment observed a nonmonotonic energy dependence of the proton and $\Lambda$ $v_1$ slope ($dv_1/dy$), with a minimum around $\sqrt{s_{\mathrm{NN}}} \sim 10$–20 GeV \cite{STAR:2014clz,STAR:2017okv,Chen:2024aom}. This behavior has been discussed in connection with possible changes in the QCD equation of state; however, existing transport and hydrodynamic models have not provided a consistent description of the detailed energy dependence \cite{Steinheimer:2014pfa,Konchakovski:2014gda,Nara:2016phs,Nara:2016hbg}. These discrepancies highlight the need for further experimental constraints, especially in the high-baryon-density regime where baryonic interactions and mean-field effects are expected to play a dominant role.

The $\phi$ meson, the lightest strange quark–antiquark bound state ($s\bar{s}$) with a mass comparable to that of the proton, provides a unique probe of the high-baryon-density regime in relativistic heavy-ion collisions \cite{STAR:2017okv,NA49:2008goy,STAR:2008inc}. In baryon-rich matter, collective dynamics is dominated by baryonic interactions and mean-field effects, while lighter mesons are strongly affected by late-stage hadronic rescattering. Owing to its hidden strangeness, long lifetime ($\sim$46 fm/$c$)\cite{SND:2024kbi}, and relatively weak effective $\phi$–$N$ interaction in nuclear matter governing rescattering and absorption in baryonic matter \cite{Hartmann:2012ia,CLAS:2010pxs,Maeda:2007zzb}, the $\phi$ meson undergoes limited rescattering and retains enhanced sensitivity to collective dynamics established during
the dense baryonic stage. Such information is mostly inaccessible through measurements of lighter mesons, which are strongly modified by late-stage rescattering, or baryons alone, whose flow is intrinsically entangled with baryon number transport and mean-field effects \cite{STAR:2021yiu,STAR:2022sir,Buss:2011mx}.

Transport model studies indicate that at high baryon chemical potential ($\mu_B$) and near production threshold, $\phi$ mesons are predominantly produced via decays of high-mass baryon resonances \cite{HADES:2009lnd,STAR:2021hyx,Fabbietti:2015tpa,Steinheimer:2015sha,Steinheimer:2025mho}. In this regime, $\phi$ mesons are created in close association with baryons and can inherit collective motion driven by baryonic mean-field interactions, in contrast to top RHIC energies where $\phi$-meson elliptic flow is commonly attributed to partonic collectivity \cite{Mohanty:2009tz,STAR:2015gge, Nasim:2014jla}. Measurements of $\phi$-meson flow at low collision energies therefore provide direct sensitivity to baryon-resonance dynamics in dense QCD matter. Following their production, the propagation of $\phi$ mesons in the dense medium is governed by their weak and medium-dependent $\phi$–$N$ interaction cross section \cite{CLAS:2010pxs,Steinheimer:2025mho}. Together with their longer lifetime, this allows $\phi$ mesons to preserve information on collective dynamics developed during the dense baryonic stage. Consequently, systematic measurements of $\phi$-meson directed flow as a function of beam energy offer new experimental sensitivity to baryon-resonance, vector-meson transport and in-medium dynamics in the high-$\mu_B$ region \cite{Rapp:2014hha,Hartmann:2012ia,Maeda:2007zzb,CLAS:2010pxs}.

In this letter, we report the $\phi$-meson rapidity-odd component of directed flow ($v_1$) in Au+Au mid-central collisions at center-of-mass energies $\sqrt{s_{\mathrm{NN}}}=3.0$, 3.2, 3.5, 3.9, 4.5, and 7.7 GeV. The $\sqrt{s_{\mathrm{NN}}}=3.0$--4.5 GeV data were collected with the STAR detector during the 2019--2021 RHIC Beam Energy Scan Fixed-Target (FXT) program, while the $\sqrt{s_{\mathrm{NN}}}=7.7$ GeV collider-energy result follows the procedure described in Ref.~\cite{STAR:2023bzh}. In the FXT configuration, a gold beam impinges on a gold target positioned at the entrance of the Time Projection Chamber (TPC) \cite{Anderson:2003ur}. Events are required to have a primary vertex within 2 cm of the target along the beam direction and within 2 cm radially to suppress background interactions with the beam pipe. Minimum-bias events are triggered using  signals from the Time-of-Flight (TOF), Beam-Beam Counter (BBC), Event Plane Detector (EPD) and Vertex Position Detector (VPD) detectors \cite{Llope:2012zz, Whitten:2008zz}.

Collision centrality is determined from charged-particle multiplicities measured by the TPC, combined with a Monte Carlo (MC) Glauber model simulation \cite{Miller:2007ri,STAR:2009sxc}. Pileup events, arising from the long TPC drift time relative to the beam bunch spacing, are suppressed by correlating the TPC multiplicity with the multiplicity of tracks matched to the Time-of-Flight (TOF) detector. In this paper, pseudorapidity $\eta$ is defined in the laboratory frame, while rapidity $y$ is given in the center-of-mass frame unless otherwise specified. The directed flow is measured with respect to the first-order Event Plane (EP) reconstructed using the EPD. For the $\sqrt{s_{\mathrm{NN}}}=7.7$ GeV collider-energy data, the event-plane reconstruction follows the procedure described in Ref.~\cite{STAR:2023bzh}. For the FXT data, the EP is reconstructed using the east side of the EPD, covering the pseudorapidity range $-5.3<\eta<-2.6$~\cite{Adams:2019fpo}. Owing to the asymmetric phase-space acceptance in the FXT geometry, the event-plane resolution is evaluated using the three-subevent method \cite{Poskanzer:1998yz,STAR:2021yiu}. In this procedure, two independent EPD subevents defined on the east side of the EPD in the pseudorapidity ranges $-5.3 < \eta < -3.3$ and $-3.3 < \eta < -2.9$ are used together with a TPC-based subevent at midrapidity ($-1 < \eta < 0$), providing a large separation in $\eta$ between the flow measurement region and the event plane determination. The resulting first-order event-plane resolution ranges from 55\% to 75\% for $\sqrt{s_{\mathrm{NN}}} = 3.0$--4.5 GeV. A centrality range of 10--40\% is selected to ensure both good event-plane resolution and sufficient $\phi$-meson yields.

\begin{figure}[ht]
\centering
   \includegraphics[trim=60 0 60 0, clip, scale=0.45]{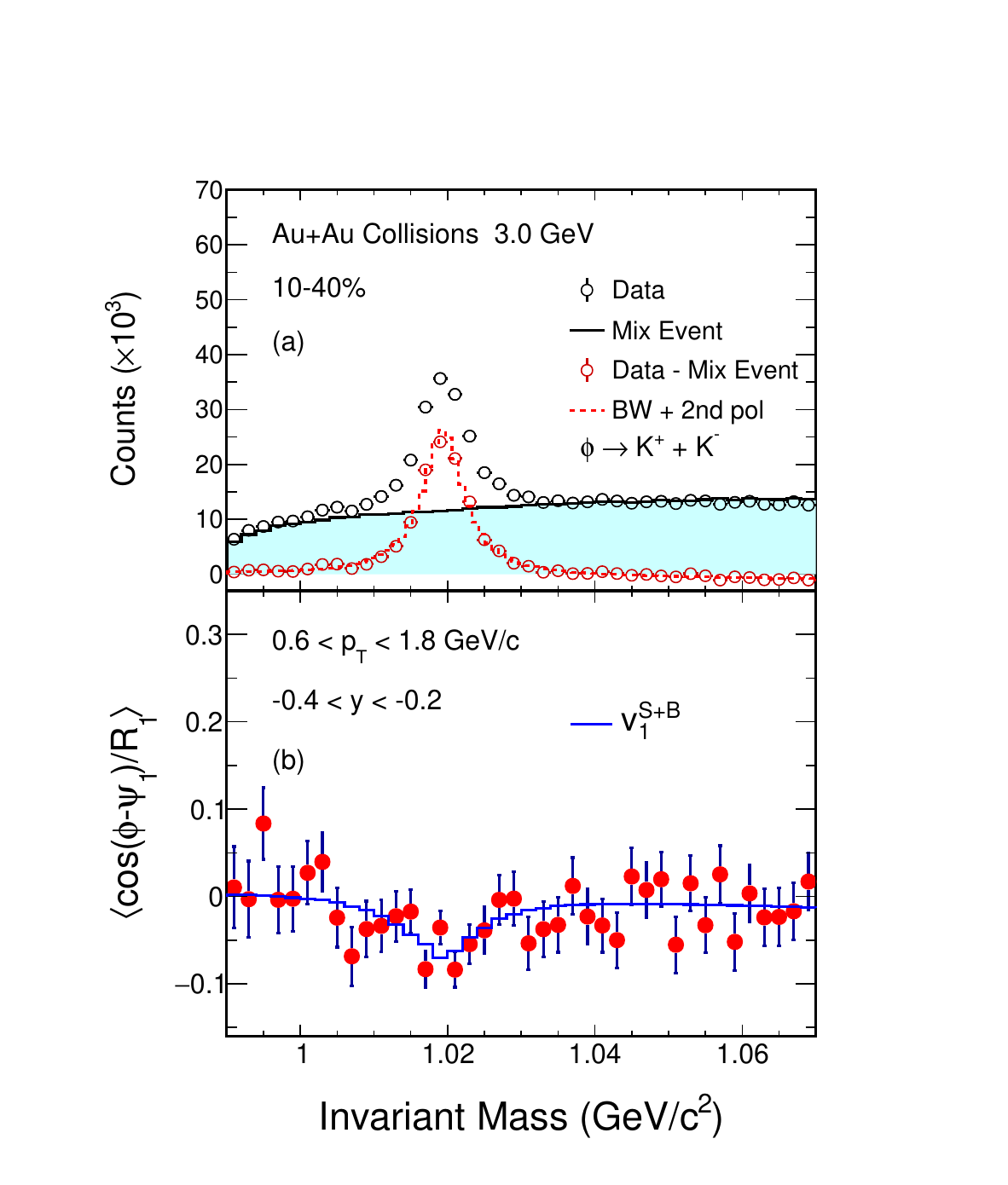}
   \caption{(a) Invariant-mass distribution of $K^{+}K^{-}$ pairs used for $\phi$-meson reconstruction in $\sqrt{s_{\mathrm{NN}}} = 3.0$~GeV Au+Au collisions (10--40\% centrality). (b) Directed flow $v_{1}$ as a function of invariant mass in the same kinematic region. The fit illustrates the separation of signal and background contributions to $v_{1}$ and validates the extraction of the $\phi$-meson directed flow.}
   \label{fig:invmass}
\end{figure}

The $\phi$ mesons are reconstructed via the decay channel $\phi \rightarrow K^+K^-$. Charged tracks are identified using combined specific energy loss ($dE/dx$) measurements from the TPC and time-of-flight information from the TOF and eTOF detectors \cite{Shao:2005iu,Xu:2008th,STAR:2002eio}. 
The invariant-mass distributions of $K^+K^-$ pairs are constructed in the center-of-mass frame as shown in Fig.~\ref{fig:invmass}(a), with combinatorial background estimated using the mixed-event (ME) technique, where $K^+$ and $K^-$ from different events with similar characteristics (e.g., centrality and event-plane angle) are paired. The background-subtracted distributions (red solid circles) are fitted with a Breit–Wigner function for the signal and a linear term for the residual background.

The $\phi$-meson directed flow is obtained using the invariant-mass method \cite{Masui:2012zh},
\begin{align}
\label{equ:mass}
    v_1^{S+B}(M) &= \left\langle \frac{\cos(\phi - \Psi_1)}{R_1} \right\rangle \\ \nonumber
    &= v_1^{S} \frac{S}{S+B}(M) + v_1^{B}(M) \frac{B}{S+B}(M),
\end{align}
where $S$ and $B$ denote the signal and background yields extracted from the invariant-mass distribution, respectively, while $v_1^{S}$ and $v_1^{B}$ represent the corresponding directed-flow components. The background $v_1^B(M)$ is parameterized as a first-order polynomial. Alternative parameterizations were also tested, and the corresponding variations are included in the systematic uncertainty. The final $\phi$-meson $v_1$ in each rapidity ($y$) bin is extracted from the fitted $v_1^S$. Following the standard convention \cite{E895:2000maf}, $v_1$ is assigned a negative sign for $y<0$. Figure~\ref{fig:invmass}(b) presents the reconstructed $v_1^{S+B}$ as a function of invariant mass and fitted with Eq.~\ref{equ:mass} in the corresponding transverse momentum ($p_T$)--rapidity ($y$) region.

\begin{figure}[ht]
\includegraphics[scale=0.44]{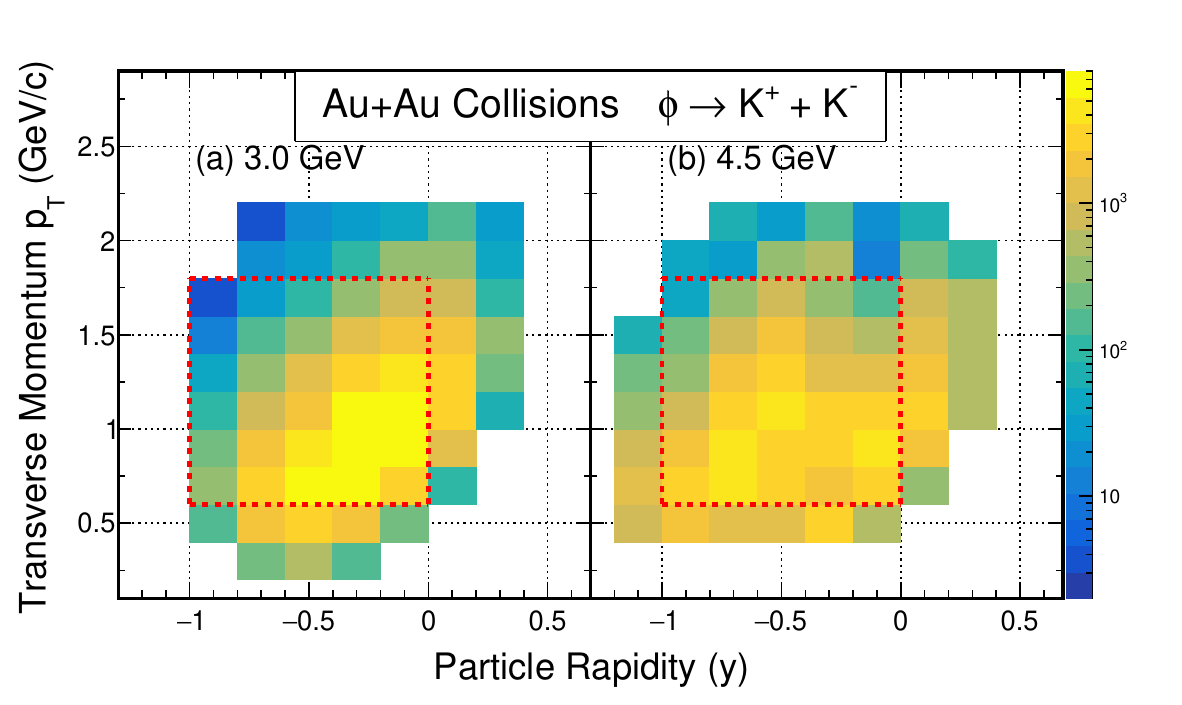}
\caption{Transverse momentum $p_T$ versus rapidity $y$ acceptance of reconstructed $\phi$-meson candidates in the center-of-mass frame for $\sqrt{s_{\mathrm{NN}}} = 3.0$~GeV (a) and 4.5~GeV (b) Au+Au collisions. The red dashed rectangles indicate the kinematic region used for the directed-flow analysis.}
\label{fig:accp}
\end{figure}

Figure~\ref{fig:accp} shows the $p_T$-$y$ acceptance of reconstructed $\phi$ candidates in the center-of-mass frame at $\sqrt{s_{\mathrm{NN}}}$ = 3 GeV (a) and 4.5 GeV (b). The red dashed boxes indicate the acceptance region used for the directed-flow analysis. Table~\ref{tab1} summarizes the acceptance windows for the particles of interest including $\phi$ as well as $K_{S}^{0}$, $p$ and $\Lambda$ for comparisons. The $p_T$ selection for $\phi$ mesons is determined by the detector acceptance, and reasonable variations of the lower $p_T$ bound (e.g., 0.4–0.5 GeV/$c$) were verified not to affect the physics conclusions. For the $p_T$-integrated $v_1$ measurements, it is essential to account for the reconstruction efficiency. The reconstruction efficiencies, including tracking and particle identification, are evaluated as functions of transverse momentum and rapidity using standard simulation-based techniques, with additional corrections derived directly from data \cite{STAR:2019bjj,STAR:2017sal,STAR:2018zdy}.

\begin{table}[htb]
    \centering
    \caption{$p_{\mathrm{T}}$-$y$ acceptance windows of $\phi$, $K_S^0$ mesons and $p$, $\Lambda$ baryons \cite{STAR:2025twg,STAR:2021yiu,STAR:2020dav} used for directed flow analysis.}    
    \begin{adjustbox}{width=0.3\textwidth}
    \begin{tabular}{lcc} \hline \hline
        Particle           & \hspace{0.5cm}$p_{\mathrm{T}}$ ({GeV/\it{c}})  & $y$ \\ \hline
        $\phi$             & (0.6, 1.8)   & (-1.0, 0.0)    \\
        $K_S^0$            & (0.4, 1.6)   & (-1.0, 0.0)    \\
        $p$                & (0.4, 2)   & (-1.0, 0.0)   \\   
        $\Lambda$          & (0.4, 2)   & (-1.0, 0.0)    \\ \hline \hline
    \end{tabular}    
    \label{tab1}
\end{adjustbox}
\end{table}

Systematic uncertainties on the midrapidity $v_1$ slope, $dv_1/dy|_{y=0}$, are evaluated by varying the track selection criteria, particle-identification requirements, the $v_1$ extraction procedure, and the reference event plane used in the event-plane resolution correction. The resulting variations are treated as independent sources and are summed in quadrature to obtain the total systematic uncertainty, as summarized in Table~\ref{tab2}. To suppress artificial contributions from statistical fluctuations in the systematic evaluation, the Barlow criterion is applied to all variations \cite{Barlow:2002yb}. Among the considered sources, the dominant systematic uncertainties are associated with the event-plane resolution correction and the track quality criteria, at the level of a few percent. Additional stability studies, including variations of the invariant-mass
fit range, background parameterization, invariant-mass binning, and an independent event-plane $dN/d(\phi - \Psi_{1})$ extraction method, yield consistent $dv_1/dy$ values within statistical and systematic uncertainties. Rapidity bins with very limited acceptance or low signal significance were checked separately and verified not to alter the extracted slopes within uncertainties.

\begin{table}[ht]
\centering
\caption{Sources of systematic uncertainties ($\%$) for the midrapidity slope $dv_1/dy|_{y=0}$ of $\phi$.}
\begin{adjustbox}{width=0.3\textwidth}
\begin{tabular}{lccccc}
\hline\hline
 & \multicolumn{5}{c}{Energy (GeV)} \\
\cline{2-6}
Source (\%) & 3.0 & 3.2 & 3.5 & 3.9 & 4.5 \\
\hline
Track quality    & 1.6 & 5.5 & 3.0 & 1.5 & 5.4 \\
PID              & 0.0 & 1.8 & 3.0 & 7.4 & 4.7 \\
$v_1$ extraction & 0.1 & 0.8 & 2.6 & 0.0 & 0.9 \\
EP resolution    & 1.6 & 1.8 & 2.8 & 4.5 & 9.1 \\
Total            & 2.2 & 6.1 & 5.7 & 8.8 & 11.6 \\
\hline\hline
\end{tabular}
\end{adjustbox}
\label{tab2}
\end{table}

\begin{figure*}[ht]
\includegraphics[scale=0.75]{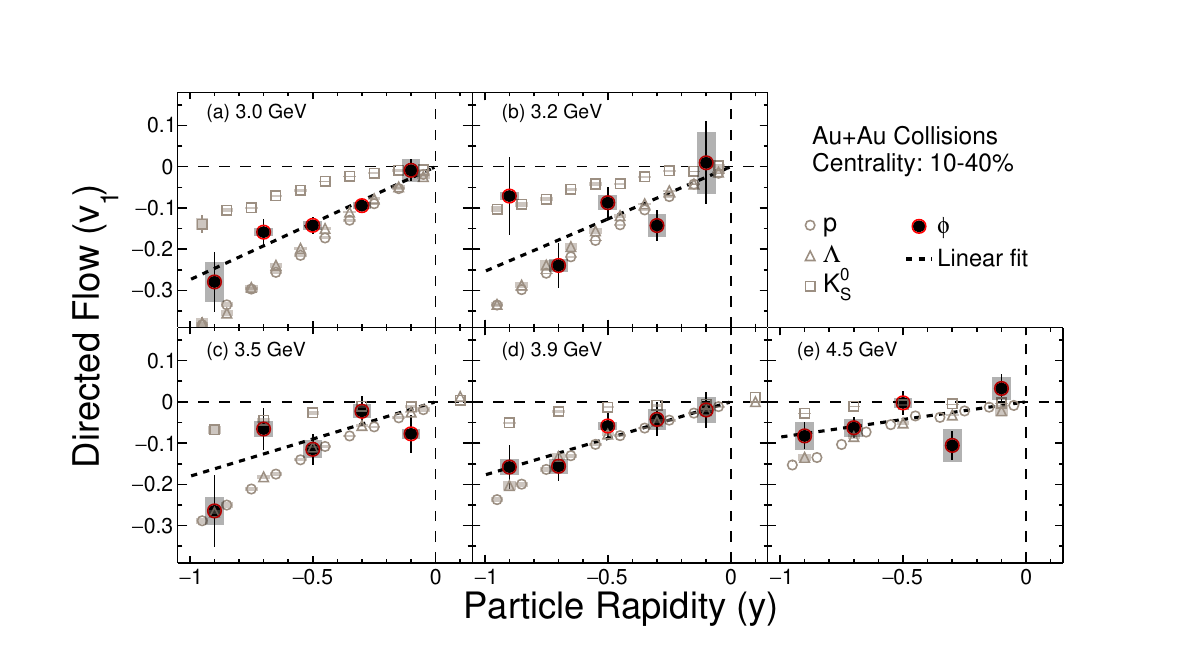}
\caption{Directed flow $v_{1}$ as a function of rapidity $y$ for $\phi$ mesons, protons ($p$), $\Lambda$ baryons, and $K^{0}_{S}$ mesons in 10--40\% Au+Au collisions at $\sqrt{s_{\mathrm{NN}}} = 3.0$--$4.5$~GeV. The solid black dashed lines represent first-order polynomial fits used to extract the midrapidity slope $dv_{1}/dy$. 
Statistical and systematic uncertainties are indicated by bars and shaded bands, respectively
}
\label{fig:v1y}
\end{figure*}


The directed flow $v_1(y)$ of $\phi$ mesons in 10–40\% mid-central Au+Au collisions at $\sqrt{s_{\mathrm{NN}}}$ = 3.0, 3.2, 3.5, 3.9, and 4.5 GeV is presented in Fig. \ref{fig:v1y}. For comparison, the $v_1(y)$ distributions of protons ($p$), $\Lambda$ baryons, and $K^0_S$ mesons \cite{STAR:2025twg,STAR:2021yiu,STAR:2020dav} are also shown with open markers. The $\phi$-meson $v_1(y)$ is similar in magnitude and sign to that of protons and $\Lambda$ baryons, but significantly larger than that of $K^0_S$. 
With increasing collision energy, the magnitude of $v_1$ for all particle species decreases.

\begin{figure}[ht]
\includegraphics[scale=0.45]{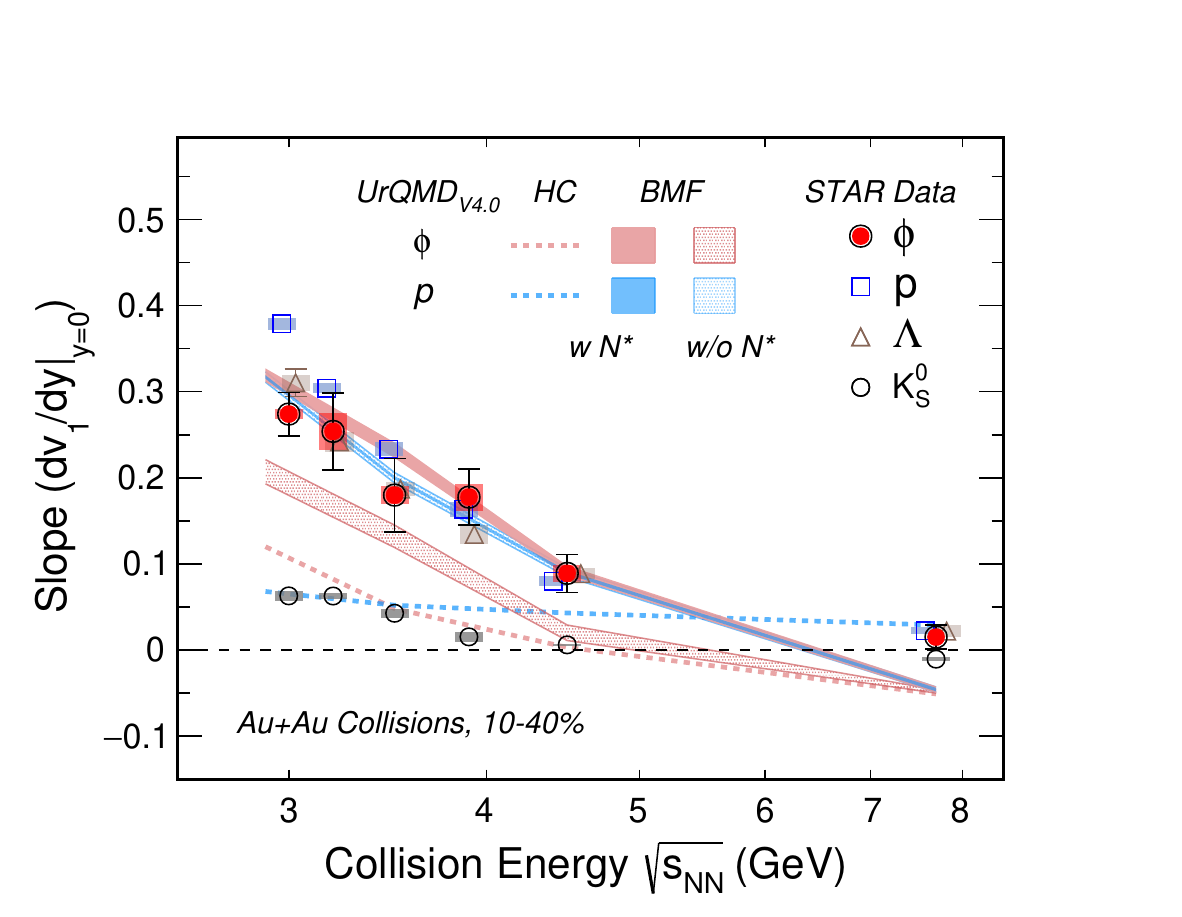}
\caption{Collision-energy dependence of the midrapidity $v_1$ slope $dv_{1}/dy$ for $\phi$ mesons, $K^{0}_{S}$ mesons, protons, and $\Lambda$ baryons in 10--40\% Au+Au collisions.
Statistical and systematic uncertainties are indicated by bars and shaded bands, respectively. Curves show UrQMD v4.0 calculations with BMF interactions plus high-mass baryon resonances (BMF+N$^{*}$), compared to calculations in the pure hadronic cascade mode without mean-fields (HC). }
\label{fig:v1slope}
\end{figure}

The midrapidity slopes $dv_1/dy|_{y=0}$ of $\phi$ mesons are extracted using linear fits to the measured $v_1(y)$ distribution in the rapidity interval $-1<y<0$. Varying the fitting window to $-1<y<-0.2$ and $-0.8<y<0$ yields consistent results within statistical uncertainties and does not contribute to the systematic uncertainty.
Fig. ~\ref{fig:v1slope} shows the collision-energy dependence of $dv_1/dy$ for $K^0_S$, $\phi$, protons, and $\Lambda$ baryons in 10–40\% Au+Au collisions at RHIC. The $v_1$ slopes of protons, $\Lambda$ baryons, and $K^0_S$ mesons are taken from previously published STAR FXT measurements \cite{STAR:2025twg,STAR:2021yiu,STAR:2020dav}, including collider data at $7.7$ GeV \cite{STAR:2014clz,STAR:2017okv}. The results at $\sqrt{s_{\mathrm{NN}}}=4.5$ GeV for $\Lambda$ and $K^0_S$ are obtained using the same analysis procedure as in this work.
For protons, $\Lambda$, and $K^0_S$, a non-linear function of the form $v_1(y) = ay + by^3$ is typically used to extract the $v_1$ slopes due to their non-linear behavior over a broad rapidity range. We have verified that the local slope extracted around midrapidity is insensitive to the choice of fitting function within the rapidity interval used in this analysis. Owing to the limited statistical precision of the $\phi$-meson data, a linear function constrained to pass through the origin ($v_1=0$ at $y=0$ by symmetry) is adopted to extract the $\phi$-meson $v_1$ slope.

A striking feature of the present results is that the $\phi$-meson directed flow is comparable in magnitude to those of protons and $\Lambda$ baryons, despite its mesonic nature. The observed similarity between the $\phi$-meson and baryon directed flow suggests that the $\phi$ meson retains sensitivity to collective baryon-driven dynamics in the high-baryon-density region.

For comparison, hadronic transport model calculations from UrQMD v4.0 \cite{Steinheimer:2014pfa,Steinheimer:2015sha}, which include contributions from high-mass baryon resonances, are also shown. When both the high-mass baryon resonances ($N^{*}$) and the baryonic mean-field (BMF) potential are included, the model reproduces the observed energy dependence of $dv_1/dy$ for protons and $\phi$ mesons reasonably well over the range $3.0 \leq \sqrt{s_{\mathrm{NN}}} \leq 4.5$ GeV.

The comparison between the measured $\phi$-mesons $v_{1}$ and UrQMD calculations indicates that BMF interactions play a central role in generating the directed flow of baryons in the high baryon density region \cite{STAR:2021yiu}. Although the direct impact of the baryonic mean field on mesons is generally expected to be weaker, the comparison with transport calculations suggests that the $\phi$-meson $v_1$ is sensitive to the presence of the BMF. In calculations without the BMF potential (hadronic cascade mode), the magnitude of $dv_1/dy$ for $\phi$ mesons is significantly underestimated, by up to a factor of three at the lowest collision energies, similar to the behavior observed for protons.

As discussed in Ref. \cite{Steinheimer:2015sha,Steinheimer:2025mho}, the agreement between the data and UrQMD calculations suggests that excitation and decay of high-mass baryon resonances at high baryon density,
\begin{align}
N+N \rightarrow N^{*}+N \rightarrow N+\phi+N ,
\end{align}
provides a natural mechanism for generating the observed $\phi$-meson directed flow. In the UrQMD implementation used for this comparison, high-mass baryon resonances contributing to this channel--such as $N^*(1990)$, $N^*(2080)$, $N^*(2190)$, $N^*(2220)$, and
$N^*(2250)$--are included. Through such intermediate baryonic states, $\phi$ mesons are produced in association with baryons and can inherit the collective motion of the baryonic medium, resulting in a $v_1$ magnitude comparable to that of protons and $\Lambda$ baryons. This production scenario is also consistent with previously observed enhancements in the yield ratios $N(\phi)/N(K^{-})$ and $N(\phi)/N(\Xi^{-})$ in 3 GeV Au+Au collisions \cite{STAR:2021hyx}. 
When BMF interactions are included but high-mass $N^{*}$ resonances are absent, UrQMD calculations fail to reproduce the measured $\phi$-meson directed flow, indicating that such resonances play an important role in describing both the enhanced $\phi$ yield and the baryon-like $\phi$-meson $v_1$ in the high-baryon-density region.

Within current hadronic transport frameworks, the involvement of high-mass baryon resonances offers a consistent interpretation of both the production and collective dynamics of $\phi$ mesons in baryon-rich matter. Further theoretical refinements may help to clarify the detailed composition of these resonance contributions, while the present data indicate that such resonances play a significant role in shaping the observed $\phi$-meson directed flow. At higher beam energies, including the $\sqrt{s_{\rm NN}}=7.7$ GeV point shown in Fig.~4, both UrQMD configurations underestimate the measured $\phi$-meson directed flow, indicating limitations of the present hadronic transport implementation in the transition from the high-baryon-density regime toward lower $\mu_B$ conditions. We note that a corresponding UrQMD comparison for $K^0_S$ is not shown in Fig.~4; nevertheless, the much smaller $K^0_S$ directed-flow magnitude over the same energy range suggests a distinct meson-species dependence of the collective dynamics.

For $\phi$-meson production at high baryon density, we observe an enhanced $\phi/K^-$ ratio \cite{STAR:2021hyx} and a strong collectivity similar to that of protons and $\Lambda$ baryons. Other transport model calculations, besides UrQMD, have also offered partial explanations. PHQMD \cite{Song:2022jcj}, employing an alternative implementation of the baryonic mean field, can reproduce the $\phi/K^-$ ratio. JAM2 \cite{Nara:1999dz,Nara:2021fuu} calculations reproduce the $v_1$ of protons and $\Lambda$ baryons, and preliminary results suggest a qualitatively consistent description of the $\phi$ $v_1$.

The present observation thus opens a new window to explore baryon resonance dynamics and vector-meson transport in dense baryonic matter, and provides experimental sensitivity to the in-medium propagation of $\phi$ mesons, including the effective $\phi$–$N$ interaction strength, as suggested in Ref. \cite{Steinheimer:2025mho}.

In summary, we report a large directed flow of $\phi$ mesons in Au+Au collisions at $\sqrt{s_{\mathrm{NN}}}=3.0$--4.5 GeV from the STAR FXT program, together with the $\sqrt{s_{\mathrm{NN}}}=7.7$ GeV collider-energy result included in the collision-energy dependence. The magnitude of the $\phi$-meson $v_{1}$ is comparable to that of protons and $\Lambda$ baryons, and significantly larger than that of $K^{0}_{S}$ mesons, indicating that $\phi$ mesons participate in the collective dynamics of baryon-rich matter.

Comparisons with hadronic transport calculations suggest that the measured $\phi$-meson directed flow is consistent with production via high-mass baryon resonances and collective baryonic motion governed by mean-field interactions. Owing to its long lifetime and relatively weak effective $\phi$--$N$
interaction, the $\phi$ meson can preserve sensitivity to collective dynamics established during the dense baryonic stage of the collision. The observed strong $\phi$-meson directed flow thus provides new experimental sensitivity to baryon-resonance production and strange-hadron transport in the high-baryon-density region.

We thank the RHIC Operations Group and SDCC at BNL, the NERSC Center at LBNL, and the Open Science Grid consortium for providing resources and support.  This work was supported in part by the Office of Nuclear Physics within the U.S. DOE Office of Science, the U.S. National Science Foundation, National Natural Science Foundation of China, Chinese Academy of Science, the Ministry of Science and Technology of China and the Chinese Ministry of Education, NSTC Taipei, the National Research Foundation of Korea, Czech Science Foundation and Ministry of Education, Youth and Sports of the Czech Republic, Hungarian National Research, Development and Innovation Office, New National Excellency Programme of the Hungarian Ministry of Human Capacities, Department of Atomic Energy and Department of Science and Technology of the Government of India, the National Science Centre and WUT ID-UB of Poland, German Bundesministerium f\"ur Bildung, Wissenschaft, Forschungand Technologie (BMBF), Helmholtz Association, Ministry of Education, Culture, Sports, Science, and Technology (MEXT), Japan Society for the Promotion of Science (JSPS), and Agencia Nacional de Investigacion y Desarrollo de Chile (ANID), Chile.    

\bibliographystyle{apsrev4-2}
\bibliography{reference.bib}

\end{document}